\documentclass[11pt,english,aps]{revtex4}
\usepackage[T1]{fontenc}
\usepackage[latin1]{inputenc}
\usepackage{geometry}
\usepackage{graphicx}
\usepackage{amssymb}

\makeatletter

\usepackage{babel}
\makeatother
\begin{document}
\begin{flushright}November 2005 \end{flushright}

\begin{center}\textbf{\Large Universality and Quark Masses of the
Staggered Fermion Action }\end{center}{\Large \par}

\author{Anna Hasenfratz}

\email{anna@eotvos.colorado.edu}

\thanks{}

\affiliation{Department of Physics, University of Colorado, Boulder, CO-80304-0390}

\begin{abstract}
Staggered fermions with 4 tastes are expected to describe 4-flavor
QCD in the continuum limit, therefore at finite lattice spacing the
staggered determinant should be equivalent to an SU(4) flavor-symmetric
system up to lattice artifacts. This equivalence is the starting point
of the 4th root trick used to reduce the number of fermion flavors
and provides the only consistent description of 2 or 1 flavor systems.
In this paper we argue that the quark mass of the underlying flavor
symmetric theory differs from the staggered mass by an additive term
due to the taste breaking of the staggered action. The relation is
the same for 2 and 1 flavor fermions. This additive mass shift implies
that at finite lattice spacing staggered simulations correspond to
heavier quark masses than indicated by the staggered Goldstone pion
and that staggered fermions cannot reach the chiral limit at finite
lattice spacing. 
\end{abstract}
\maketitle

\section{Note Added }

The justification of the 4th root trick is essential for the large
scale high precision simulations carried out in recent years using
2+1 flavors staggered sea quarks. The validity of the procedure depends
on the properties of the staggered action. In this paper I assumed
that the unrooted 4-taste staggered action differs from a flavor degenerate
4-flavor action only in local pure gauge terms and investigated the
consequences of this assumption. 

Since the original publication of this manuscript there is growing
evidence indicating that this assumption is incorrect \cite{Bernard:2006xx,Hasenfratz:2006xx}%
\footnote{Many of these recent developments were discussed in the {}``Workshop
on the 4th root of the straggered determinant'' at INT, Seattle,
March 2006.%
}. In a recent paper the author and Hoffmann \cite{Hasenfratz:2006xx}
presented numerical evidence in the 2 dimensional Schwinger model
that indicates that while the rooted staggered action might not be
local at finite lattice spacing, in the continuum limit the non-local
terms become irrelevant and the rooted action can describe chiral
fermions with arbitrary physical mass. The same paper also justifies
the ad-hoc assumption of this manuscript about the mass shift between
the physically matched staggered and overlap fermions. I would like
to emphasize that this mass offset is a lattice artifact. For further
details the reader is referred to the publication \cite{Hasenfratz:2006xx}
and references there.

\section{Introduction}

Staggered fermion lattice actions have four non-degenerate flavors
or tastes. There is little doubt that in the continuum limit they
correspond to QCD with four degenerate flavors. Consider a renormalization
group transformation that respects gauge and chiral symmetry, with
a fixed point on the $m=0$ and $g^{2}=0$ critical surface that represents
4-flavor continuum QCD. Along the renormalized trajectory of the RG
transformation the lattice actions describe the same continuum system;
these actions presumably have Ginsparg-Wilson chiral, but at least
SU(4) flavor symmetry. The renormalization group transformation should
move the staggered lattice action toward this renormalized trajectory,
and since RG transformations change the action only in irrelevant,
cut-off level terms, the SU(4) flavor symmetric action along the renormalized
trajectory will differ from the original staggered one only in irrelevant,
cut-off level terms. 

In this paper we investigate this underlying SU(4) flavor symmetric
action at finite lattice spacing. We argue that its bare fermion mass
must differ from the staggered mass not only in a multiplicative but
in an additive mass renormalization factor as well. This additive
term is related to the taste breaking of the staggered action and
vanishes only in the continuum limit. An immediate consequence of
this observation is that the staggered Goldstone pion can be significantly
lighter that the flavor symmetric pions, leading to incorrect mass
estimates at finite lattice spacing. We support our argument with
data form the 2-dimensional Schwinger model.

\section{Decomposition of the Staggered Determinant}

The 4-taste staggered determinant describes 4-flavor QCD in the continuum
limit if its determinant can be decomposed as \begin{equation}
{\rm {det}}(D_{{\rm {st}}})={\rm {det}(D_{{\rm {s4}}}){\rm {det}(T)}}\:,\label{eq:Basic}\end{equation}
where $D_{{\rm {st}}}$ is the staggered Dirac operator, $D_{{\rm {s4}}}$
is an SU(4) flavor symmetric lattice operator, and the operator $T$
gives cut-off level corrections only. If one could rigorously show
that the action $\tilde{S}=-{\rm {Tr\, ln}(T)}$ was a local operator
in the continuum limit, Eq.\ref{eq:Basic} would validate the 4th
root procedure used in reducing the number of flavors in staggered
simulations \cite{Adams:2004mf,Durr:2004as,Durr:2004ta,Shamir:2004zc,Bernard:2005gf}.
We are not concerned about this issue right now, so all we assume
is that $T$ does not change the physical predictions.

Both $D_{{\rm {st}}}$ and $D_{{\rm {s4}}}$ has one relevant operator
corresponding to the fermion mass. (The relevant direction in the
gauge field is governed by the gauge action which is not specified
in Eq.\ref{eq:Basic}.) We will denote the corresponding bare masses
by $m_{q}$ and $m_{{\rm s4}}$. The renormalized quark masses differ
from the bare ones by multiplicative renormalization factors.

Eq.\ref{eq:Basic} comes from a RG based operator relation, it has
to be valid on all, or at least most, gauge configurations. Consider
a topologically non-trivial configuration at finite lattice spacing.
In the chiral $m_{{\rm {s4}}}=0$ limit $D_{{\rm {s4}}}$ has 4 degenerate
zero modes, ${\rm {det}(D_{{\rm {s4}}})=0}$. On the other hand $D_{{\rm {st}}}$
has no exact zero modes at any finite lattice spacing, ${\rm {det}(D_{{\rm {st}}})\ne0}$
even when $m_{q}=0$. The determinant of the cut-off level term ${\rm T}$
has to be finite, so Eq.\ref{eq:Basic} cannot be valid if both $m_{q}=0$
and $m_{{\rm {s4}}}=0$. Yet Eq.\ref{eq:Basic} is necessary if 4-taste
staggered fermions are in the same universality class as 4-flavor
continuum QCD. The only way to satisfy the equation is to allow $m_{{\rm {s4}}}$
to remain finite when the staggered quark mass vanishes, i.e. \begin{equation}
m_{{\rm {s4}}}=a^{2}\Delta+Z\, m_{q}.\label{eq:additive}\end{equation}
 The additive correction is $O(a^{2})$ but present at any finite
lattice spacing. 

Eq.\ref{eq:Basic} suggests two different ways to look at 4-taste
staggered fermions. The first is the usual staggered interpretation:
the action describes four fermionic flavors but due to lattice artifacts
the flavor symmetry is broken, only a remnant U(1) symmetry survives.
At the massless $m_{q}=0$ limit the system has one Goldstone pion,
the other pseudoscalars have $O(a^{2})$ masses. Only in the continuum
limit do we recover the full SU(4)xSU(4) flavor symmetry. The second
interpretation is based on the right hand side of Eq.\ref{eq:Basic}.
It states that the theory has 4 degenerate fermions, it differs from
a Ginsparg-Wilson action only in cut-off terms, so in the $m_{{\rm {s4}}}=0$
limit we expect 15 massless pions. The two interpretations are compatible
but their corresponding fermionic fields are very different. They
have different lattice artifacts and their predictions can also differ
by $O(a^{2})$ effects, but in the continuum limit they are identical.
For a consistent field theory one has to use the same action for the
valence as for the sea quarks. For 4-taste staggered fermions we could
use either the staggered action or the SU(4) flavor symmetric action
in the valence sector.

The relation between the staggered and flavor symmetric quark masses
is the same for 1-flavor staggered fermions defined by the 4th root
of the staggered determinant, assuming the 4th root does not introduce
non-universal, non-local terms. There is an important difference though
between the 4-taste and 1-flavor cases. The 4th root procedure is
correct only if we consider the SU(4) flavor symmetric interpretation,
so for a consistent field theoretical description one has to use this
4th root operator in the valence quark sector as well. When the standard
staggered operator is used in the valence sector we have to deal with
partial-quenching/mixed actions effects, and also a mismatch between
the staggered and flavor symmetric quark masses. Problems due to partial-quenching/mixed
actions clearly show up when one attempts to calculate the pseudoscalar
singlet \cite{Gregory:2005me,Prelovsek:2005rf}. One should note that
these inconsistencies are not related to the 4th root determinant
being non-local, they simply show the inconsistencies between the
sea and valence quark actions. 

Staggered chiral perturbation theory \cite{Lee:1999zx,Aubin:2003mg,Aubin:2003uc}
models the 4th root by hand correcting the number of fermions in the
sea quark loops. That approach assumes that the 4th root of the staggered
Dirac operator is a local theory. We know that this is not the case
\cite{Bunk:2004br}. For a consistent perturbative treatment one should
consider staggered valence quarks on the background of Ginsparg-Wilson
quarks whose mass is adjusted/fitted appropriately. Mixed action simulations
has to match the valence and sea quark masses. Again, this matching
should be done to the SU(4) flavor symmetric masses.

A finite additive mass implies that with staggered fermions one cannot
reach the chiral limit at any finite lattice spacing. This might explain
the contradictions of the finite temperature scaling behavior of $N_{f}=2$
simulations. For these systems the O(4) critical point is beyond the
reach of the simulations, it lies at negative quark mass, though it
likely influences the scaling behavior at positive mass values as
well. At the same time there is an O(2) invariant critical point at
$m_{q}=0$ that mainly effects the fermionic observables. The combination
of these two nearby critical points can make the scaling of the system
very complicated.

\section{Numerical Support}

In a series of papers Durr and Hoelbling investigated the 2-dimensional
Schwinger model with 1 and 2 flavors both with staggered and overlap
fermions \cite{Durr:2003xs,Durr:2004ta}. Here we consider their results
for the topological susceptibility. 

\begin{figure}
\includegraphics[%
  width=6.5in]{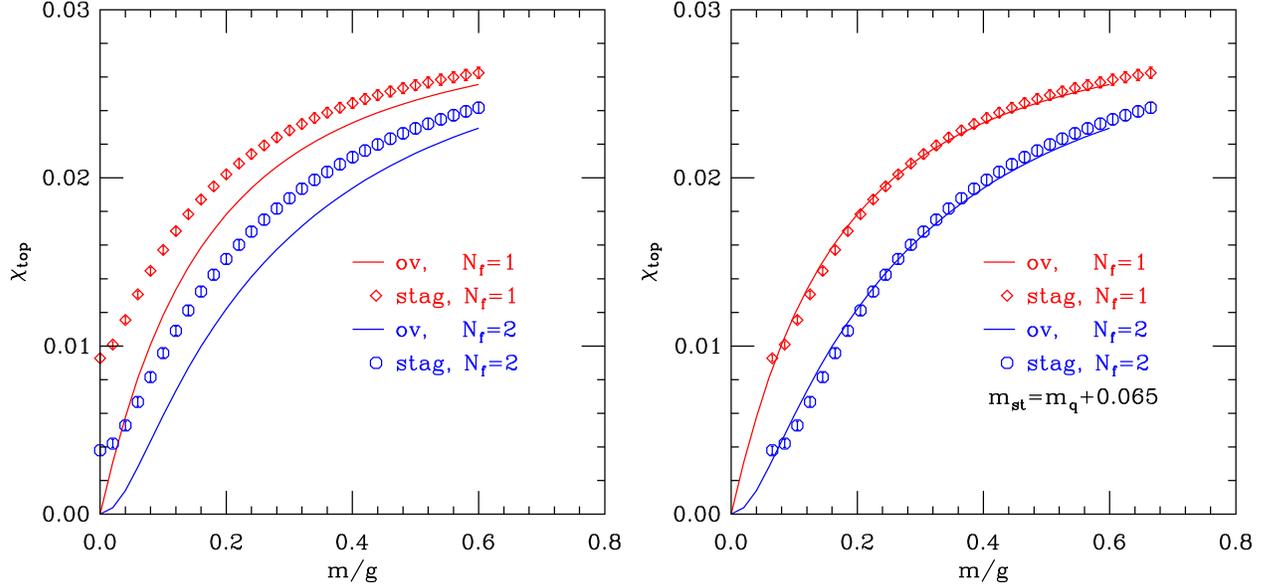}

\caption{The topological susceptibility in the 2D Schwinger model with $N_{f}=1$
and 2 flavors with overlap and staggered dynamical fermions. The left
panel is from Ref.\cite{Durr:2003xs}. The right panel is the same
data replotted with the quark mass for the staggered fermions shifted
as $m=m_{q}+0.065$. \label{cap:The-topological-susceptibility} }
\end{figure}

In Ref.\cite{Durr:2003xs} the topological susceptibility was measured
through the index of the massless overlap operator, so any deviation
between the staggered and overlap action is due to the difference
between the dynamical configurations. Chiral perturbation theory predicts
that the topological susceptibility vanishes in the chiral limit.
The solid curves in Fig. \ref{cap:The-topological-susceptibility}
show the susceptibility both on $N_{f}=1$ and 2 overlap configurations.
Both curves follow the expected behavior. The diamonds and octagons
on the left panel show the result on staggered dynamical configurations
\footnote{The author is indebted to S. Durr for providing the data for the plot.%
}. The susceptibility remains finite even at the $m_{q}\to0$ limit
indicating that the underlying chiral theory has a finite mass. Re-plotting
the staggered data with a mass shift $m_{q}\to m_{q}+0.065$ (right
panel of Fig. \ref{cap:The-topological-susceptibility}) brings the
overlap and staggered curves together. The agreement is not perfect
but very good. Including a multiplicative $Z$ factor like in Eq.\ref{eq:additive}
as $m=1.05m_{q}+0.055$ makes the agreement excellent everywhere but
at the lowest $m_{q}=0$ point. The fact that both the $N_{f}=1$
and 2 data set requires the same mass shift indicates that both are
governed by the same chiral action with $m\approx m_{q}+0.065$. This
observation supports the expectation that the square root action of
$N_{f}=1$ is a correct 1-flavor action. The data in Fig. \ref{cap:The-topological-susceptibility}
corresponds to unsmeared fermions. Even one level of APE smearing
reduces the difference between the staggered and overlap actions,
the required mass shift becomes $m-m_{q}\lesssim0.01$. Of course
the taste violation is also similarly reduced after a smearing step.

There were other unexpected differences between the staggered and
overlap results in Refs. \cite{Durr:2003xs,Durr:2004ta}, mainly with
the scalar condensate. It is likely that those too will be resolved
by observing the mass shift between the two actions.

One would like to estimate the mass shift for the 4D Asqtad action.
Measurements of the topological susceptibility clearly indicate a
mass shift \cite{Bernard:2003gq}. Staggered chiral perturbation theory
predicts that the topological susceptibility scales with the taste-singlet
pion\cite{Billeter:2004wx}, suggesting that the additive mass is
governed by the heaviest of the staggered pions.

\section{Conclusion}

We argued that if 4-taste staggered fermions describe 4-flavor QCD
in the continuum limit, at any finite lattice spacing the fermion
mass of the underlying chiral 4-flavor theory differs from the mass
of the staggered action not only by a multiplicative but by an additive
term as well. The same relation holds between the masses for 1 and
2-flavor systems. 

This additive mass is a lattice artifact and will go away in the continuum
limit. At finite lattice spacing however staggered measurements can
underestimate the light hadron masses if those are extracted from
the lowest mass staggered operators. The mass shift can have strong
effects on finite temperature simulations and can explain the large
lattice artifacts observed in the topological susceptibility as well.
The additive mass should be taken into account in any simulation using
mixed actions.

\section{Acknowledgments}

This work was supported by the US Department of Energy. I am indebted
to T. DeGrand for numerous discussions. I would like to thank S. Sharpe
and S. Durr for extensive correspondence, P. Hasenfratz and F. Niedermayer
for helpful questions and comments.

\bibliographystyle{apsrev}
\bibliography{lattice,lattice}

\end{document}